\documentclass[12pt]{article}
\usepackage{graphicx}

\begin{document}

\title{\textbf{Momentum dependence of the symmetry potential and its influence on nuclear reactions}}

\author{Zhao-Qing Feng \footnote{Corresponding author. Tel. +86 931 4969215. \newline \emph{E-mail address:} fengzhq@impcas.ac.cn (Z.-Q. Feng)}}
\date{}
\maketitle

\begin{center}
\small \emph{Institute of Modern Physics, Chinese Academy of
Sciences, Lanzhou 730000, People's Republic of China}
\end{center}

\textbf{Abstract}
\par
A Skyrme-type momentum-dependent nucleon-nucleon force distinguishing isospin effect is parameterized and further implemented in the Lanzhou Quantum Molecular Dynamics (LQMD) model for the first time, which leads to a splitting of nucleon effective mass in nuclear matter. Based on the isospin- and momentum-dependent transport model, we investigate the influence of momentum-dependent symmetry potential on several isospin-sensitive observables in heavy-ion collisions. It is found that symmetry potentials with and without the momentum dependence but corresponding to the same density dependence of the symmetry energy result in different distributions of the observables. The mid-rapidity neutron/proton ratios at high transverse momenta and the excitation functions of the total $\pi^{-}/\pi^{+}$ and $K^{0}/K^{+}$ yields are particularly sensitive to the momentum dependence of the symmetry potential.

\bigskip

Heavy-ion collisions induced by neutron-rich nuclei at intermediate and relativistic energies in terrestrial laboratories are a useful tool to extract the information of
nuclear equation of state (EoS) of isospin asymmetric nuclear matter, which is poorly known for the high-density neutron-rich matter but has an important application in astrophysics, such as the structure of neutron star, the cooling of protoneutron stars, the nucleosynthesis during supernova explosion of massive stars etc \cite{St05}. The EoS of nuclear matter is usually expressed through the energy per nucleon as $E(\rho,\delta)=E(\rho,\delta=0)+E_{\textrm{sym}}(\rho)\delta^{2}+\textsc{O}(\delta^{2})$
in terms of baryon density $\rho=\rho_{n}+\rho_{p}$, relative neutron excess $\delta=(\rho_{n}-\rho_{p})/(\rho_{n}+\rho_{p})$,
energy per nucleon in a symmetric nuclear matter $E(\rho,\delta=0)$ and bulk nuclear symmetry energy
$E_{\textrm{sym}}=\frac{1}{2}\frac{\partial^{2}E(\rho,\delta)}{\partial\delta^{2}}\mid_{\delta=0}$. Based on several complementary analysis of available experimental data associated with transport models, a symmetry energy of $E_{\textrm{sym}}(\rho)\approx 31.6(\rho/\rho_{0})^{\gamma}$ MeV with $\gamma=0.69-1.05$ was extracted for densities between 0.1$\rho_{0}$ and 1.2$\rho_{0}$ \cite{Li08}. However, predictions for high-density symmetry energies based on various microscopical or phenomenological many-body theories diverge widely \cite{Di03,Ch05,Li06}. More realistic approach to extract the information of the $E_{\textrm{sym}}(\rho)$ is the comparison between transport model calculations and experimental data. For that, the reliable input potentials including the isovector (symmetry potential) and isoscalar parts in the transport models are very necessary for precisely predicting some promising observables.

The momentum dependence of the isoscalar potential leads to the same nucleon effective mass (Landau mass) for neutrons and protons in nuclear matter and has been widely studied in transport models for heavy-ion collisions. The effective mass splitting of neutrons and protons results from the momentum-dependent symmetry potential, which has been implemented in the one-body transport models, such as the isospin Boltzmann-Uehling-Uhlenbeck (IBUU04) model \cite{Li04} and the stochastic mean-field (SMF) model \cite{To99,Ba05}. In this work, a Skyrme-type momentum dependent nucleon-nucleon (NN) force distinguishing protons and neutrons is parameterized and then included in a N-body approach (LQMD model) for the first time. Furthermore, effects of the momentum dependence in heavy-ion collisions are investigated. In particular, its influence on the isospin sensitive observables to extract the high-density symmetry energy is discussed.

The LQMD model has been successfully applied to treat the dynamics in heavy-ion fusion reactions near Coulomb barrier and also to describe the capture of two heavy colliding nuclides to form a superheavy nucleus \cite{Fe05,Fe08}. Further improvements of the LQMD model were performed in order to investigate the dynamics of pion and strangeness productions in heavy-ion collisions and also to extract the information of isospin asymmetric EoS at supra-saturation densities \cite{Fe09,Fe10a,Fe10b,Fe10c}. In the previous versions, we only considered the scalar part of the momentum-dependent interaction and the density-dependent symmetry potential. We have included the resonances ($\Delta$(1232), N*(1440), N*(1535)), hyperons ($\Lambda$, $\Sigma$) and mesons ($\pi$, $K$, $\eta$) in hadron-hadron collisions and the decays of resonances for treating heavy-ion collisions in the region of 1A GeV energies. The time evolutions of the baryons (nucleons and resonances) and mesons in the system under the self-consistently generated mean-field are governed by Hamilton's equations of motion, which read as
\begin{eqnarray}
\dot{\mathbf{p}}_{i}=-\frac{\partial H}{\partial\mathbf{r}_{i}},
\quad \dot{\mathbf{r}}_{i}=\frac{\partial H}{\partial\mathbf{p}_{i}}.
\end{eqnarray}
We only consider the Coulomb interaction for charged hyperons. The Hamiltonian of baryons consists of the relativistic energy, the effective interaction potential and
the momentum dependent part as follows:
\begin{equation}
H_{B}=\sum_{i}\sqrt{\textbf{p}_{i}^{2}+m_{i}^{2}}+U_{int}+U_{mom}.
\end{equation}
Here the $\textbf{p}_{i}$ and $m_{i}$ represent the momentum and the
mass of the baryons.

The effective interaction potential is composed of the Coulomb
interaction and the local interaction
\begin{equation}
U_{int}=U_{Coul}+U_{loc}.
\end{equation}
The Coulomb interaction potential is written as
\begin{equation}
U_{Coul}=\frac{1}{2}\sum_{i,j,j\neq
i}\frac{e_{i}e_{j}}{r_{ij}}erf(r_{ij}/\sqrt{4L})
\end{equation}
where the $e_{j}$ is the charged number including protons and
charged resonances. The $r_{ij}=|\mathbf{r}_{i}-\mathbf{r}_{j}|$ is
the relative distance of two charged particles.

The local interaction potential is derived directly from the Skyrme
energy-density functional and expressed as
\begin{equation}
U_{loc}=\int V_{loc}(\rho(\mathbf{r}))d\mathbf{r}.
\end{equation}
The local potential energy-density functional reads
\begin{eqnarray}
V_{loc}(\rho)=&&  \frac{\alpha}{2}\frac{\rho^{2}}{\rho_{0}}+
\frac{\beta}{1+\gamma}\frac{\rho^{1+\gamma}}{\rho_{0}^{\gamma}}+
\frac{g_{sur}}{2\rho_{0}}(\nabla\rho)^{2}+\frac{g_{sur}^{iso}}{2\rho_{0}}
[\nabla(\rho_{n}-\rho_{p})]^{2}      \nonumber \\
&&  +E_{sym}^{loc}(\rho)\rho\delta^{2}+g_{\tau}\rho^{8/3}/\rho_{0}^{5/3},
\end{eqnarray}
where the $\rho_{n}$, $\rho_{p}$ and $\rho=\rho_{n}+\rho_{p}$ are
the neutron, proton and total densities, respectively, and the
$\delta=(\rho_{n}-\rho_{p})/(\rho_{n}+\rho_{p})$ is the isospin
asymmetry. The coefficients $\alpha$, $\beta$, $\gamma$, $g_{sur}$,
$g_{sur}^{iso}$, $g_{\tau}$ are related to the Skyrme parameters
$t_{0}, t_{1}, t_{2}, t_{3}$ and $x_{0}, x_{1}, x_{2}, x_{3}$ \cite{Fe08}. The $E_{sym}^{loc}$ is the local part of the symmetry energy, which can be adjusted to mimic predictions calculated by microscopical or phenomenological many-body theories and has two-type forms as follows:
\begin{equation}
E_{sym}^{loc}(\rho)=\frac{1}{2}C_{sym}(\rho/\rho_{0})^{\gamma_{s}},
\end{equation}
and
\begin{equation}
E_{sym}^{loc}(\rho)=a_{sym}(\rho/\rho_{0})+b_{sym}(\rho/\rho_{0})^{2}.
\end{equation}
The parameters $C_{sym}$, $a_{sym}$ and $b_{sym}$ are taken as 52.5 MeV, 43 MeV, -16.75 MeV and 38 MeV, 37.7 MeV, -18.7 MeV for the cases with and without momentum-dependent interactions, respectively. The values of $\gamma_{s}$=0.5, 1., 2. correspond to the soft, linear and hard symmetry energy, respectively, and the Eq. (8) gives a supersoft symmetry energy, which cover the largely uncertain of nuclear symmetry energy, particularly at the supra-saturation densities.

We have taken the same Skyrme-type form for the momentum-dependent potential in the Hamiltonian as in Ref. \cite{Ai87} but distinguishing isospin effect, which is expressed as
\begin{equation}
U_{mom}=\frac{1}{2\rho_{0}}\sum_{i,j,j\neq i}\sum_{\tau,\tau'}C_{\tau,\tau'}\delta_{\tau,\tau_{i}}\delta_{\tau',\tau_{j}} \int\int\int d\textbf{p}d\textbf{p}'d\textbf{r} f_{i}(\textbf{r},\textbf{p},t)[\ln(\epsilon(\textbf{p}-\textbf{p}')^{2}+1)]^{2}f_{j}(\textbf{r},\textbf{p}',t).
\end{equation}
The term is also given from the energy-density functional in nuclear matter,
\begin{equation}
U_{mom}=\frac{1}{2\rho_{0}}\sum_{\tau,\tau'}C_{\tau,\tau'} \int\int\int d\textbf{p}d\textbf{p}'d\textbf{r} f_{\tau}(\textbf{r},\textbf{p})[\ln(\epsilon(\textbf{p}-\textbf{p}')^{2}+1)]^{2}f_{\tau'}(\textbf{r},\textbf{p}').
\end{equation}
Here $C_{\tau,\tau}=C_{mom}(1+x)$, $C_{\tau,\tau'}=C_{mom}(1-x)$ ($\tau\neq\tau'$) and the isospin symbols $\tau$($\tau'$) represent proton or neutron. The sign of $x$ determines different mass splitting of proton and neutron in nuclear medium, e.g. positive signs corresponding to the case of $m^{\ast}_{n}<m^{\ast}_{p}$. The parameters $C_{mom}$ and $\epsilon$ were determined by fitting the real part of optical potential as a function of incident energy from the proton-nucleus elastic scattering data, which determine the nucleon effective mass in isospin symmetric nuclear matter. In the calculation, we take the values of 1.76 MeV, 500 c$^{2}$/GeV$^{2}$ and -0.65 for the $C_{mom}$, $\epsilon$ and $x$, respectively, which result in the effective mass $m^{\ast}/m$=0.75 in nuclear medium at saturation density for symmetric nuclear matter. For the cold nuclear matter, we have the phase-space density $f_{\tau}(\textbf{r},\textbf{p})=\rho_{\tau}(\textbf{r})\Theta(p_{F}(\tau)-|\textbf{p}|)/(4\pi p_{F}^{3}(\tau)/3)$ with the Fermi momentum $p_{F}(\tau)=\hbar(3\pi^{2}\rho_{\tau})^{1/3}$. Implementing the phase-space distribution into Eq. (10), we get the contribution of the momentum dependence to the symmetry energy $E_{sym}^{mom}(\rho)$. Therefore, the symmetry energy per nucleon in the LQMD model is composed of three parts, namely the kinetic energy, the local part and the momentum dependence of the potential energy as
\begin{equation}
E_{sym}(\rho)=\frac{1}{3}\frac{\hbar^{2}}{2m}\left(\frac{3}{2}\pi^{2}\rho\right)^{2/3}+E_{sym}^{loc}(\rho)+E_{sym}^{mom}(\rho).
\end{equation}
Figure 1 is a comparison of different stiffness of nuclear symmetry energy after inclusion of the momentum-dependent interactions. One can see that all cases cross at saturation density with the value of 31.5 MeV. The local part of the symmetry energy can be adjusted to reflect the uncertain behavior of the symmetry energy at sub- and supra-normal densities. The contributions of the local and momentum-dependent interactions of the potential part in the symmetry energy is shown in Fig. 2. The momentum dependence has a positive contribution to the symmetry energy in the mass splitting of $m^{\ast}_{n}<m^{\ast}_{p}$, which is opposite for the case of $m^{\ast}_{n}>m^{\ast}_{p}$. The local part of the symmetry energy is adjusted to get the same density dependence of the symmetry energy for a given stiffness with increasing density.


Combined Eq. (6) and Eq. (10), we get a density, isospin and momentum-dependent single-particle potential in nuclear matter as follows:
\begin{eqnarray}
U_{\tau}(\rho,\delta,\textbf{p})=&&  \alpha\frac{\rho}{\rho_{0}}+\beta\frac{\rho^{\gamma}}{\rho_{0}^{\gamma}}+\frac{8}{3}g_{\tau}\rho^{5/3}/\rho_{0}^{5/3}+
E_{sym}^{loc}(\rho)\delta^{2}+\frac{\partial E_{sym}^{loc}(\rho)}{\partial\rho}\rho\delta^{2}+
E_{sym}^{loc}(\rho)\rho\frac{\partial\delta^{2}}{\partial\rho_{\tau}}       \nonumber \\
&&  +\frac{1}{\rho_{0}}C_{\tau,\tau} \int d\textbf{p}' f_{\tau}(\textbf{r},\textbf{p})[\ln(\epsilon(\textbf{p}-\textbf{p}')^{2}+1)]^{2}      \nonumber \\
&&  +\frac{1}{\rho_{0}}C_{\tau,\tau'} \int d\textbf{p}' f_{\tau'}(\textbf{r},\textbf{p})[\ln(\epsilon(\textbf{p}-\textbf{p}')^{2}+1)]^{2}.
\end{eqnarray}
Here $\tau\neq\tau'$, $\partial\delta^{2}/\partial\rho_{n}=4\delta\rho_{p}/\rho^{2}$ and $\partial\delta^{2}/\partial\rho_{p}=-4\delta\rho_{n}/\rho^{2}$. The effective (Landau) mass in nuclear matter is calculated through the potential as $m_{\tau}^{\ast}=m_{\tau}/ \left(1+\frac{m_{\tau}}{|\textbf{p}|}|\frac{dU_{\tau}}{d\textbf{p}}|\right)$ with the free mass $m_{\tau}$ at Fermi momentum $\textbf{p}=\textbf{p}_{F}$. Therefore, the nucleon effective mass only depends on the momentum-dependent interactions. The mass splitting of protons and neutrons in nuclear matter as functions of density ($\delta$=0.2) and isospin asymmetry ($\delta=(\rho_{n}-\rho_{p})/(\rho_{n}+\rho_{p})$) at saturation density is shown in Fig. 3. The left windows are the cases of $m^{\ast}_{n}<m^{\ast}_{p}$ with the parameter $x$=0.65 in the coefficient $C_{\tau,\tau'}$ and the right panels are $m^{\ast}_{n}>m^{\ast}_{p}$. The two sorts of the mass splitting can be chosen in the LQMD calculations. In accordance with the Lane potential \cite{La62}, the symmetry potential can be evaluated from the single-nucleon potential $U_{sym}(\rho,\textbf{p})=(U_{n}(\rho,\delta,\textbf{p})-U_{p}(\rho,\delta,\textbf{p}))/2\delta$.


A hard core scattering in two-particle collisions is assumed in the simulation of the collision processes by Monte Carlo procedures, in which the scattering of two particles is determined by a geometrical minimum distance criterion $d\leq\sqrt{0.1\sigma_{tot}/\pi}$ fm weighted by the Pauli blocking of the final states \cite{Ai91,Be88}. Here, the total cross section $\sigma_{tot}$ in mb is the sum of the elastic and all inelastic cross sections. The probability reaching a channel in a collision is calculated by its contribution of the channel cross section to the total cross section as $P_{ch}=\sigma_{ch}/\sigma_{tot}$. The choice of the channel is done randomly by the weight of the probability. The primary products in nucleon-nucleon (NN) collisions in the region of 1A GeV energies are the resonances $\Delta$(1232), $N^{\ast}$(1440), $N^{\ast}$(1535) and the pions. We have included the reaction channels as follows:
\begin{eqnarray}
&& NN \leftrightarrow N\triangle, \quad  NN \leftrightarrow NN^{\ast}, \quad  NN
\leftrightarrow \triangle\triangle,  \nonumber \\
&& \Delta \leftrightarrow N\pi,  N^{\ast} \leftrightarrow N\pi,  NN \rightarrow NN\pi (s-state),  N^{\ast}(1535) \rightarrow N\eta.
\end{eqnarray}
At the considered energies, there are mostly $\Delta$ resonances which disintegrate into a $\pi$ and a nucleon in the evolutions. However, the $N^{\ast}$ yet gives considerable contribution to the energetic pion yields. The energy and momentum-dependent decay widths are used in the calculation \cite{Fe09} for the $\Delta$(1232) and $N^{\ast}$(1440) resonances. We have taken a constant width $\Gamma$=150 MeV for the $N^{\ast}$(1535) decay. The strangeness is created by inelastic hadron-hadron collisions \cite{Fe10c,Fe11}. We included the channels as follows:
\begin{eqnarray}
&& BB \rightarrow BYK,  BB \rightarrow BBK\overline{K},  B\pi \rightarrow YK,  B\pi \rightarrow NK\overline{K}, \nonumber \\
&& Y\pi \rightarrow B\overline{K}, \quad  B\overline{K} \rightarrow Y\pi, \quad YN \rightarrow \overline{K}NN.
\end{eqnarray}
Here the B strands for (N, $\triangle$, N$^{\ast}$) and Y($\Lambda$, $\Sigma$), K(K$^{0}$, K$^{+}$) and $\overline{K}$($\overline{K^{0}}$, K$^{-}$). The elastic scattering between strangeness and baryons are considered through the channels $KB \rightarrow KB$, $YB \rightarrow YB$ and $\overline{K}B \rightarrow \overline{K}B$. The evolutions of mesons ($\pi$, $K$, $\eta$) are also determined by the Hamiltonian in which the Coulomb interaction and the in-medium potential were considered in the model \cite{Fe10b,Fe11}.

To check the influence of the momentum dependence of the symmetry potential on reaction dynamics, we calculated the transverse momentum distributions of the ratios of neutrons over protons in the mid-rapidity domain and the excitation functions of charged pion yields as shown in Fig. 4. Inclusion of the momentum-dependent interaction in the symmetry potential reduces the n/p yields at high transverse momentum owing to its negative contribution to the symmetry energy at the case of $m^{\ast}_{n}>m^{\ast}_{p}$ in nuclear medium, which enforces an attractive force in neutron-neutron collisions and further increases the collision probabilities. Contrarily, the case without the isovector of the momentum-dependent interaction quickly squeezes out neutrons, in particular for the energic neutrons, which nearly appears a flat distribution at $p_{t}>$0.3 GeV/c. The conclusions are consistent with the calculations based on the IBUU04 transport model \cite{Li04}. The charged pion ratios are slightly changed by the momentum-dependent potential. Therefore, the neutron and proton transverse emission ratio in the mid-rapidity region is to be a nice probe of the isovector part of the momentum-dependent interaction.


The pions and strange particles are mainly produced in the dense hadronic matter formed in heavy-ion collisions. The $\pi^{-}/\pi^{+}$ and $K^{0}/K^{+}$ ratios can be probes to extract the high-density information of isospin asymmetric EoS. Shown in Fig. 5 is a comparison of the excitation functions of the $\pi^{-}/\pi^{+}$ and $K^{0}/K^{+}$ ratios with and without the momentum dependence of the symmetry potential in the $^{197}$Au+$^{197}$Au reaction for head-on collisions. One can see that the momentum-dependent interaction by distinguishing isospin effect reduces the $\pi^{-}/\pi^{+}$ and $K^{0}/K^{+}$ yields, which is more pronounced close to the threshold energies of meson production. The results are caused from the fact that the interaction enhances the energic neutron-neutron collisions. Furthermore, the $\pi^{-}$ and $K^{0}$ is produced through the channels $nn \rightarrow p\Delta^{-}$, $\Delta^{-} \rightarrow n\pi^{-}$ and $nn \rightarrow n\Lambda K^{0}$.


In summary, influence of the momentum dependence of the symmetry potential on isospin sensitive observables in heavy-ion collisions is investigated by using an isospin- and momentum-dependent transport model (LQMD). It is found that the momentum dependence of the symmetry potential plays an important role on nucleon transverse emissions and the ratios of $\pi^{-}/\pi^{+}$ and $K^{0}/K^{+}$, which are also as promising probes of high-density symmetry energy. To precisely constrain the density dependence of the nuclear symmetry energy, one firstly needs to get the accurate information of the density- and momentum-dependent symmetry potential. Further experimental data of determining the momentum-dependent symmetry potential are very necessary to investigate accurately the dense neutron-rich matter. The updated LQMD model would to be a useful tool to predict the density dependence of the symmetry energy from heavy-ion collisions, in particular at supra-normal densities, and also to analyze experimental data for constraining the symmetry energy.

\textbf{Acknowledgements}

This work was supported by the National Natural Science Foundation of China under Grant No. 10805061; the Special Foundation of the
President Fund and the West Doctoral Project of Chinese Academy of Sciences.

\newpage
\begin{figure}
\begin{center}
{\includegraphics*[width=0.8\textwidth]{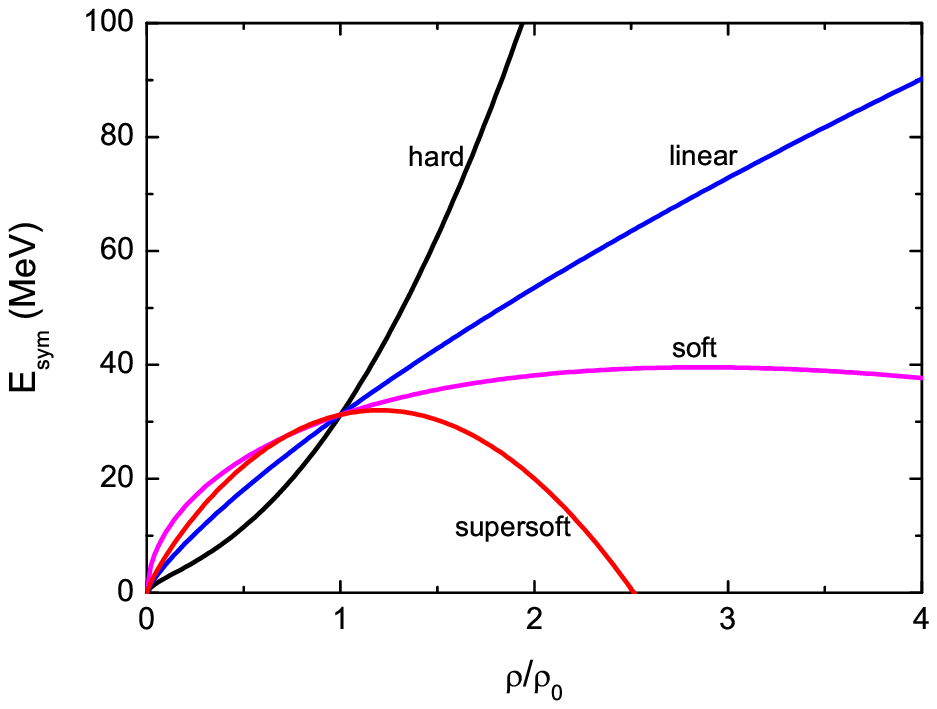}}
\end{center}
\caption{Density dependence of the nuclear symmetry energy for the cases of supersoft, soft, linear and hard trends.}
\end{figure}

\begin{figure}
\begin{center}
{\includegraphics*[width=0.8\textwidth]{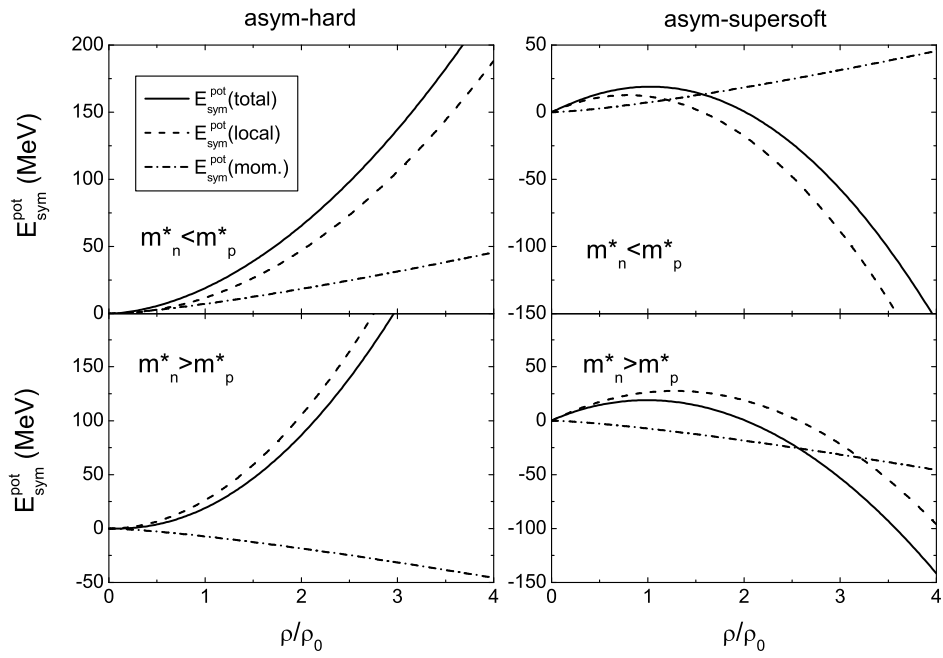}}
\end{center}
\caption{Density dependence of the potential part of nuclear symmetry energy at different mass splitting for hard (left panel) and supersoft (right panel) symmetry energies.}
\end{figure}

\begin{figure}
\begin{center}
{\includegraphics*[width=0.8\textwidth]{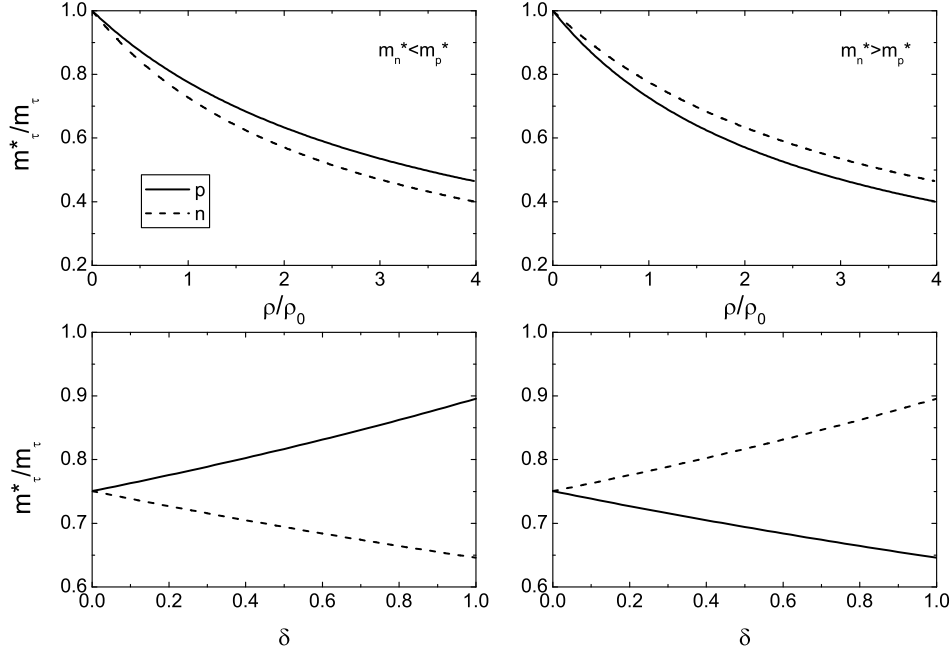}}
\end{center}
\caption{Nucleon effective mass normalized by vacuum mass as functions of density ($\delta=0.2$) and isospin asymmetry ($\rho=\rho_{0}$).}
\end{figure}

\begin{figure}
\begin{center}
{\includegraphics*[width=0.8\textwidth]{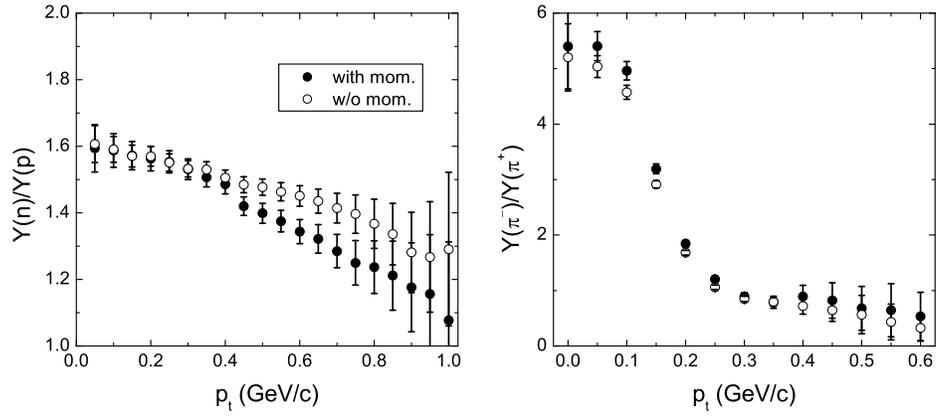}}
\end{center}
\caption{Transverse momentum distributions of the ratios of neutron/proton within the rapidity bin $|y/y_{proj}|<$0.3 and $\pi^{-}/\pi^{+}$ in central $^{124}$Sn+$^{124}$Sn collisions at incident energy 400A MeV.}
\end{figure}

\begin{figure}
\begin{center}
{\includegraphics*[width=0.8\textwidth]{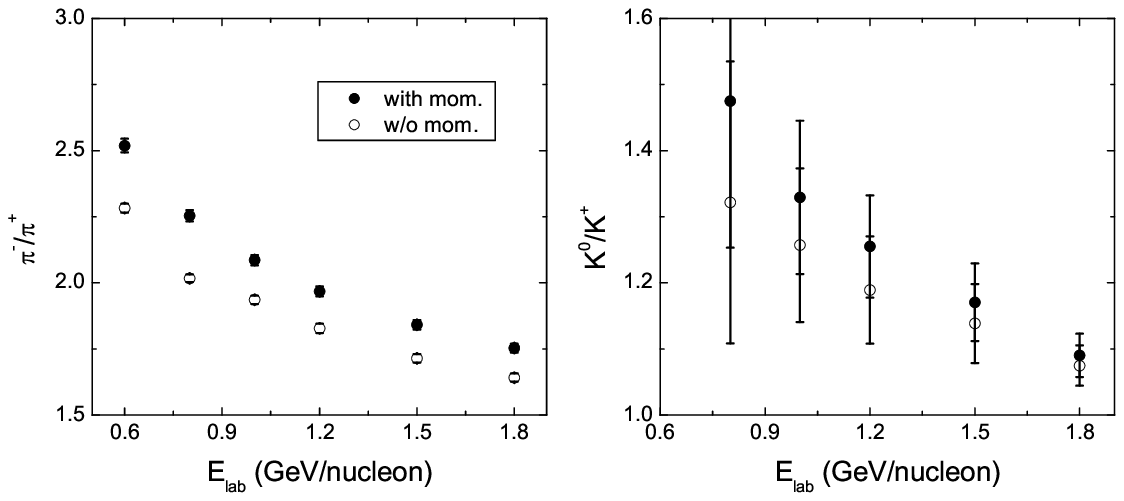}}
\end{center}
\caption{Comparison of excitation functions of the $\pi^{-}/\pi^{+}$ and $K^{0}/K^{+}$ yields for head-on collisions in the $^{197}$Au+$^{197}$Au reaction with and without the isovector part of momentum dependence.}
\end{figure}


\begin{thebibliography}{99}
\bibitem{St05} A.W. Steiner, M. Prakash, J.M. Lattimer, P. J. Ellis, Phys. Rep. 411 (2005) 325.
\bibitem{Li08} B.A. Li, L.W. Chen, C.M. Ko, Phys. Rep. 464 (2008) 113.
\bibitem{Di03} Dieperink, et al., Phys. Rev. C 68 (2003) 064307.
\bibitem{Ch05} L.W. Chen, C.M. Ko, B.A. Li, Phys. Rev. C 72 (2005) 064309; Phys. Rev. C 76 (2007) 054316.
\bibitem{Li06} Z.H. Li, et al., Phys. Rev. C 74 (2006) 047304.
\bibitem{Li04} B.A. Li, C. B. Das, S. Das Gupta, C. Gale, Phys. Rev. C 69 (2004) 011603(R); Nucl. Phys. A 735 (2004) 563.
\bibitem{To99} M. Di Toro, et al., Prog. Part. Nucl. Phys. 42 (1999) 125.
\bibitem{Ba05} V. Baran, M. Colonna, V. Greco, M. Di Toro, Phys. Rep. 410 (2005) 335.
\bibitem{Fe05} Z.Q. Feng, F.S. Zhang, G.M. Jin, X. Huang, Nucl. Phys. A 750 (2005) 232; Z.Q. Feng, et al., Chin. Phys. Lett. 22 (2005) 3040.
\bibitem{Fe08} Z.Q. Feng, G.M. Jin, F.S. Zhang, Nucl. Phys. A 802 (2008) 91; Z.Q. Feng, G.M. Jin, Phys. Rev. C 80 (2009) 037601.
\bibitem{Fe09} Z.Q. Feng, G.M. Jin, Chin. Phys. Lett. 26 (2009) 062501.
\bibitem{Fe10a} Z.Q. Feng, G.M. Jin, Phys. Lett. B 683 (2010) 140.
\bibitem{Fe10b} Z.Q. Feng, G.M. Jin, Phys. Rev. C 82 (2010) 044615.
\bibitem{Fe10c} Z.Q. Feng, G.M. Jin, Phys. Rev. C 82 (2010) 057901.
\bibitem{Ai87} J. Aichelin, A. Rosenhauer, G. Peilert, et al., Phys. Rev. Lett. 58 (1987) 1926.
\bibitem{La62} A.M. Lane, Nucl. Phys. 35 (1962) 676.
\bibitem{Ai91} J. Aichelin, Phys. Rep. 202 (1991) 233.
\bibitem{Be88} G.F. Bertsch, S. Das Gupta, Phys. Rep. 160 (1988) 190.
\bibitem{Fe11} Z.Q. Feng, arXiv:1102.4696.

\end{thebibliography}
\end{document}